# Why customers participate in social commerce activities? – A laddering analysis

**Syahida Hassan**
School of Information Management
Victoria University of Wellington
Wellington, New Zealand
Email: syahida.hassan@vuw.ac.nz

**Janet Toland,**
School of Information Management
Victoria University of Wellington
Wellington, New Zealand
Email: janet.toland@vuw.ac.nz

**Mary Tate**
School of Information Management
Victoria University of Wellington
Wellington, New Zealand
Email: mary.tate@vuw.ac.nz

## Abstract

A new phenomenon emerging within virtual communities is a blurring between social and commercial activities. This paper explores the motivations of customers who participate in social commerce, The chosen context is micro-businesses operated by members of a virtual community of Malay lifestyle bloggers. Observations were carried out and 21 participants were interviewed in order to build an understanding of the community. We used laddering techniques in order to answer the research questions, which is "what is the relationship between participation in virtual communities and their participation in social commerce?" We found that virtual community relationship was the main influential factor, and that virtual community relationship contributed to the sense of social support as well as customers' trust in social commerce.

**Keywords** Social commerce, participation, virtual community relationship, social support

## 1 Introduction

The emergence of Web 2.0 and associated features has encouraged users to create their own content and resulted in a growth in the number of virtual communities. From an e-commerce perspective, Web 2.0 has created new opportunities for reaching customers as well as promoting products. Social media technology has provided new insights for the marketing and advertising communities. This has influenced the growth of social commerce. Social commerce is a subset of e-commerce that involves using social media to assist in e-commerce transactions and activities, and also supports social interactions and user content contributions (Liang, Ho, Li and Turban, 2011).

The opportunities associated with social commerce have generated significant interest for both researchers and practitioners (Zhao et al. 2013), however to date most discussions about social commerce have come from the medium to large organization's point of view (Stephen and Toubia 2010, Stephen and Galak 2010, Culnan et al. 2010, Zhao and Benyoucef 2013). Although there are a rising number of research articles focusing on customers in social commerce (Dong-Hee 2013, Sanghyun and Park 2013, Afrasiabi and Benyoucef 2011, Ng 2013, Guo et al. 2011), very few focus on customers of micro-businesses or also known as businesses that operates on a very small scale. Brandtzaeg and Heim (2009) argue that it is necessary for IS researchers to understand the goals and personal incentives of people who use such services, as it will be of benefit to sellers in strategizing their moves. Customers' feedback on business strategy is important for company growth, this is because knowing and understanding customers is at the centre of every successful business. Thus, this study argues that obtaining feedback on customers' experience is crucial for business. As a result of focusing solely on the business side, scholars have missed some of the most important elements of social commerce, for example virtual community involvement. This limits a fuller understanding of how the sense of virtual community contributes to participation in social commerce.

Amongst the unresolved issues in social commerce is the problem of how to turn "likes" into purchases. Manish Mehta, Vice President of Social Media & Community Division of Dell in Sahota (2010) claims that if customers actually buy products through social media, rather than just liking them, it could change the nature of social media use by business. This is supported by Jarvenpaa & Tuunanen (2012) who claim that the firms are still charting their course in terms of linking customer participation in social commerce to revenue performance. Most business are still having difficulty in fully understanding why customers interact with them using social media revealing a perception gap between a business and its customers (Heller-Baird & Parasnis, 2011). A deeper understanding of the role the virtual community plays would provide more opportunities for researchers to understand how to exploit the role of the virtual community in social commerce. Thus it is important to address this gap. Although there is some research focusing on customers, it is limited to exploring the interaction amongst customers themselves, rather than between customers and the business. Therefore, we argue that there is a need to explore communities which nurture communication between sellers and customers. In addition, most social commerce studies focus on brand communities rather than on general virtual communities (Goh et al. 2012, Gummerus et al. 2012). We believe that an investigation of this topic is essential in order to fully recognize the role of the virtual community in social commerce from the perspective of the customers of micro-businesses.

This study investigates the factors that contribute to the motivation of customers to participate in social commerce. Specifically, we explore (i) the motivations of customers of micro-businesses operating in the Malay blogosphere, and (ii) how these motivations are different to customers using more traditional and established e-commerce sites.

At present, there is still debate among researchers regarding the definitions and scope of social commerce. Beach (2005), Beisel (2005), Decker (2007), Gregoriadis (2007), Pagani & Mirabello (2011), and Rubel (2006) define social commerce as a community of shoppers who share knowledge and sources on product information, which also includes assistance from trusted individuals on goods and services, sold online. This is similar to the definition by Raito (2007) who argues that social commerce is a trusted environment where customers dynamically contribute content for the referral and sale of goods or services through positive and negative feedback. However, Raito's definition is slightly different from the others as he specifies that the trusted environment in social commerce is restricted to a network of friends and family. Meanwhile, Liang et al. (2011), Marsden (2009) Turban, Lee, King, Liang & Turban (2012), and Wang & Zhang (2012) define social commerce as a subset of e-commerce that involves the use of social media to support social interaction and user contributions, which assist the online buying and selling of products and services. Gibbons (2008), Deragon (2008), Matsumoto (2009), Afrasiabi and Benyoucef, (2011) and Weaver (2010) give similar definitions whereby social commerce is defined as the use of social media in the context of e-commerce or a form of commerce mediated by social media. Lastly, according to Stephen and Toubia (2010), social commerce depicts the marketplace where the individual sellers, instead of firms; are connected with each other through social networks. Based on this definition, it is clear that Stephen and Toubia (2010) do not include customers' activities under the scope of social commerce. According to them, customers' activities using social media are referred to as Social Shopping. However, this argument has been challenged by other researchers. For example, Wang (2009) states that social shopping and social commerce are the same concept, which is a new type of e-commerce linking shopping and social networking through social media.

Based on the previous discussion, it can be seen that the term social commerce originated from the idea of knowledge sharing about goods and/or services among customers. It also shows that the scope of social commerce differs based on the type of implementation, which triggers the need to carry out further research on the different types of implementation models that exist in the social commerce market.

It has been observed that social commerce seems to be mushrooming in South East Asian countries, such as Malaysia, Singapore and Indonesia (Fletcher and Greenhill 2009). In Malaysia, social commerce has shown significant growth over the last 7 to 8 years. This may be because Malaysia has the second-highest social network penetration in Southeast Asia, with 91% of the Internet population engaged in some sort of social network (Unicef 2014).

Initially, according to Armesh, Salarzehi, Yaghoobi, and Nikbin (2010), many Malaysian customers were still reluctant to release payment card information to online merchants, fearing losing control over their bank account. However it has been observed that social commerce has changed this perception. As more businesses, including micro-businesses, notice this growing trend for online shopping in Malaysia, they are starting to use social commerce.

As a result of this phenomenon, starting from 1 July 2013, all businesses and services conducted online will have to comply with the Consumer Protection (Electronic Trade Transactions) Regulations 2012. This regulation is imposed on all online sellers including those selling online via a blogshop or social media online store, which shows that the Malaysian government is well aware of these trends. The question that arises from this phenomena is what makes the Malaysians choose social commerce over traditional e-commerce sites? Based on our initial observation, we propose that their active participation as lifestyle blogger community members is amongst the factors that influence Malaysian to choose social commerce as compared to traditional e-commerce. This paper uncovers the real reasons behind this phenomenon and untangles the relationship between virtual community participation and social commerce participation, particularly on the customers' side.

Recent studies on social commerce customers have explored the role of social support and relationship quality amongst customers in the social commerce community (Liang et al. 2011; Hajli 2015; Zhang et al. 2014), the impact of the technological environment (Zhang et al. 2014), user preferences of social features (Huang & Benyoucef (2014) and co-branding intention (Wang & Hajli, 2014). Liang et al. (2011), Hajli (2015), Zhang et al. (2014) and Wang & Hajli (2014) found that participation in social commerce is likely to be influenced by social support and relationship quality in the community. Social support is is one of the most important functions of social relationships. Social support is always intended to be helpful by the sender, thus distinguishing it from interactions which are intentionally negative (such as angry criticism, hassling, undermining) (Haeney & Israel, 2008). Liang et al. (2011), Hajli (2015), Zhang et al. (2014) and Wang & Hajli (2014) have investigated specific communities of customers such as a micro-blogging community, a Facebook community, the Renren community and the SinaWeibo community. Renren and SinaWeibo are the premier social network platforms in China. Although their study discovered with significant findings related to relationship quality and social support, as mentioned before, nonetheless the contexts chosen for their studies didn't include a consideration of the interaction between sellers and the customers.

Hajli (2015) investigated social media constructs such as forums & communities, ratings and reviews, and recommendations and referrals and the role such constructs played within social commerce environments. He found that consumers are increasingly using these constructs as a medium for social interaction. The interactions eventually lead them to become closer to each other and influences their participation. This leads to the concept of virtual community relationship, which is defined as the personal friendships developed between members of the blogosphere community. These relationships often develop through private online communication, and they sometimes move into face-to-face interactions about the common topic of interest. The relationships were based on previous interactions and experiences, with others in the virtual community environment. Koh & Kim (2003) define this as the sense of virtual community, which consists of the dimensions of membership, influence, and immersion and reflect respectively the affective, cognitive, and behavioural aspects of virtual community members. We proposed that this relationship can also develop between sellers and the customers in the same community, and that the seller-customer relationship which includes the concept of social support will have some influence on the customers' motivation to participate in social commerce activities.

## 2　Lifestyle bloggers community

This study explores the social commerce activities of a community of lifestyle bloggers. This community is a subset of the Malaysian blogosphere community. Hopkins (2011) claims that lifestyle bloggers start off their blogs using the 'personal' genre - i.e. diaristic accounts of individuals' lives. They initially develop their own community with the same circle of followers. The followers may or may not be bloggers. If they are successful in attracting readers, some then evolve into social commerce as a result of getting paid by advertisers for a proportion of advertising space and advertorials in their blogs. On top of these advertisements, they often also sell goods (eg: apparel, food, electronics etc.) and services (eg: make-up artist, event planner etc.) via their blogs targeting their readers. The past 7- 8 years has seen a significant growth in the number of micro-businesses in this bloggers lifestyle community.

The characteristic of a blog's ease of use means there are low barriers to starting up a social commerce business. An online store can be conveniently set up without any cost, merging blogging and business into one. Bloggers also often use other social media such as Facebook and Instagram for their business attracting further customers from the same community of other bloggers and readers as well as potential customers outside the community. In other words, the social commerce implementation for this community are using multiple social media platforms. At this point, for the members who venture into business, the relationship amongst bloggers and readers is no longer limited to the bloggers-readers but moving towards more of a sellers-customers relationship. This is what makes this community unique because the sellers are also members of the community.

## 3   Data collection and analysis

In an effort to answer the research questions, we have conducted interviews with 21 customers, aged 25-35 years old members, who participate actively in the lifestyle bloggers' community. The reason why the participants were chosen from the same community is because this study is trying to establish the relationship between their participation in the community and their participation in social commerce within the same community. Therefore customers from outside the community were not being considered in this research. The participants chosen were those who read blogs at least 4-5 times a week, may or may not own blog, actively communicate with other people in the community, be it with other readers or bloggers, and have experience purchasing goods and/or services from the sellers in the same community. They were selected based on observations which were conducted at an earlier stage of the research in order to understand the selected community (i.e. their activities; communication etc.). Some of the customers in this community, also play a role as a sellers. However, to avoid bias, the participants chosen for this research were those who were not engaged in selling. They were also briefed about the research in order to make them understand that the questions being asked were about their role as customers and community members.

The interviews were carried out using the laddering technique. Laddering is an in-depth interviewing technique used to develop an understanding of how consumers translate the attributes of products and services into meaningful associations with respect to self (Reynolds and Gutman 1998). It adapts means-end theory, which was originally a framework for comprehensively representing the consumer meanings that underlie product-positioning (Gengler and Reynolds 1995). According to Gengler and Reynolds (1995), the means-end framework provides a much richer understanding of how customers derive a unique perspective based on the personal meaning of their purchases. By adapting means-end-theory, the laddering technique uses a hierarchical organization of consumer perceptions and product knowledge to get to the root reason for the purchase (Reynolds and Gutman 1998; Wansik 2003) which includes (i) product's attribute (Attribute), (ii) consequences of product use (Consequence) and (iii) individuals' values/goals (Core Values).

In IS research, the laddering technique is usually conducted for user experience (UX) or human-computer interaction research. For example Bleumers et al. (2012), Zaman (2011) and Abeele et al. (2012). For this research, the technique was adapted to identify the factors that contribute to the motivation behind customers' participation in social commerce. Most previous social commerce studies have used either quantitative analysis techniques such as surveys and/or general qualitative techniques such as interviews, by selecting the laddering approach we are hoping to dig deeper into the problem by getting to the root reasons forparticipation.

In order to conduct laddering, we firstly needed to determine the attributes which will be used as a starting point. This is called the elicitation technique. We combined two elicitation techniques which were (i) to choose from a list; and (ii) free elicitation. For the first technique, participants were asked to choose from a list of potential activities in social commerce that were generated from the observation phase. Participants were asked to pick one or more activities in social commerce as the ones that they most frequently participate in. The free elicitation technique was also used to encourage them to add on their own preferences on top of the activities itemized in the list. Next, based on their answers, which were denoted as attributes; the participants were pushed up the ladder by asking "why" questions. For example: Why is it (the attribute) important to you? In this study, the questions started as follows: Why is it important for you to participate in social commerce? This stage was followed by probing questions that examined some of the consequences associated with the attributes (Wansik 2003) until the values behind each of the activities were clear.

Once the data was finalized, it was analysed using three main steps (i) content analysis, (ii) construction of implication matrix, and (iii) development of hierarchical value map. The construction of the implication matrix and hierarchical value map were carried out using Ladderux software.

Content analysis for laddering serves to reduce the raw data in order to facilitate interpretation. It consists of two steps: data reduction and categorization. Data reduction involves the coding of each interview transcript and the development of constructs (Refer Table 1). After data reduction, a categorization process is conducted in order to categorize the constructs into different dimensions. This was done by first classifying all construct from the transcription into the three basic A/C/V (Attributes/Consequences/Values) levels (Refer to column 3, Table 1). A set of summary codes was developed to reflect the ladders of A-C-V elements (Refer Table 2)

| **TRANSCRIPTS** | **CODE** | **A/C/V** |
|---|---|---|
| Activity: Purchase | Purchase | Attribute |
| Why is it important for you to purchase from other members of the community? *As I mentioned before, at first I didn't think much of purchasing stuff online. But it finally hits me that I shouldn't just simply purchase from anyone because there is lots of thing going on social media. Scammers are out there. So I go back to the basic, buy stuff from <u>someone we know, or at least we are familiar with</u>. So that is why I <u>purchase from my blogger friends</u>.* | VC Relationship | Consequences |
| What about knowing them that you think matters to you? *Because I <u>trust</u> them.* | Trust | Consequences |
| What is so important about trust? *I am <u>comfortable</u> to deal with someone I trust. It is <u>easy to communicate and discuss</u> things with them* | Perceived Convenience | Consequences |
| Why is it important to be comfortable? *So that I don't have to think about the risk because as I said before there are too many people with bad intention out there* | Risk awareness | Consequences |
| Why is it important? I mean to have the low risk? *I don't have to worry much about the deal. And that also means that there will be <u>high possibility I will return to the same sellers</u>. To think of it, yes, most of the times, I <u>bought things from the same sellers</u> when I realized there is no risk.* | Loyalty | Values |

Table 1: Coding and the development of constructs

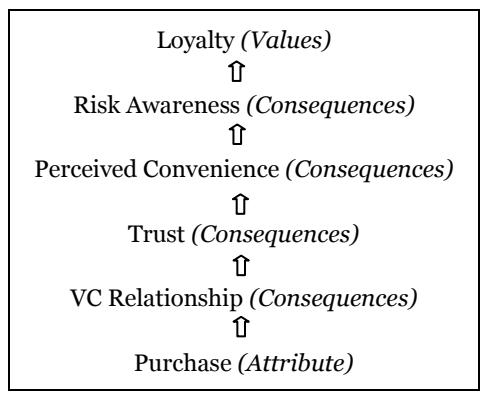

Table 2: Ladder example of A-C-V elements for Table 1 transcription.

After the ladders were constructed for each participants, the data was input using the Ladderux software which automatically transformed it into an implication matrix table and HWM. The implication matrix displays the number of times each element leads to every other element in the same ladders. In other words, it is used to summarise the connections between each attribute, consequence, and value elements. There are two types of connections that exist between elements: (i) Direct and (ii) Indirect relations. A-C-V ladders, a direct connection between two elements exists when one element has a direct link to the other element, without any intermediary elements in the same ladder. For example, in Table 3, the connection between the "Purchasing" and "Trust" is 2.12. This figure shows that there are 2 purchasing activities done directly because of the trust and there are 12 purchasing activity done because of various reason but indirectly connected to trust. The direct linkages appear on the left of the decimal (2) and indirect links on the right (12). Both direct and indirect links are shown in the implication matrix on Table 3. Note that Table 3 only displays the first 15 elements from 29 elements found in this study. Strong relationships in the study are indicated by the darker shaded boxes. The total number of ladders constructed in for this study is 46 ladders. Participants had identified 570 linkages with a total of 202 direct links and a total of 368 indirect links.

|  | 04 | 05 | 06 | 07 | 08 | 09 | 10 | 11 | 12 | 13 | 14 | 15 |
|---|---|---|---|---|---|---|---|---|---|---|---|---|
| 01 Purchasing | 18.1 | 2 | 3 | 3 | 0.4 | 0.6 | 0.3 | 0.1 | 2.12 |  | 0.3 | 0.6 |
| 02 Read Review | 3 |  |  |  |  |  |  |  | 3.3 | 0.4 | 0.3 | 0.3 |
| 03 Value Creation | 6 |  | 1 |  | 1.4 | 1.4 | 0.7 |  |  |  |  |  |
| 04 VC Relationship |  | 0.1 |  |  | 3.3 | 2.5 | 4.3 | 0.1 | 14 | 0.1 | 0.4 | 0.4 |
| 05 New Media Att. |  |  |  |  | 0.1 | 2 | 0.1 |  |  |  |  |  |
| 06 WOM |  |  |  |  |  |  |  |  | 1 |  |  |  |
| 07 Product Variety |  |  |  |  |  |  | 0.1 |  |  |  | 1 | 2 |
| 08 Social Support |  |  |  |  |  | 7.1 | 5.2 |  | 1 |  |  |  |
| 09 Social Norms |  |  |  |  |  |  | 3.3 |  | 0.1 |  |  |  |
| 10 Sense of Obligation |  |  |  |  |  |  |  |  |  |  |  | 0.1 |
| 11 Social Image |  |  |  |  |  |  |  |  | 0.1 |  | 1 |  |
| 12 Trust |  |  |  |  |  |  |  |  |  | 4 | 2.3 | 3.3 |
| 13 Product Info |  |  |  |  |  |  |  |  |  |  | 1 | 1.1 |
| 14 Decision Quality |  |  |  |  |  |  |  |  |  |  |  | 2 |
| 15 Product Quality |  |  |  |  |  |  |  |  |  |  |  |  |

Table 3: The first 15 elements in the Implication Matrix

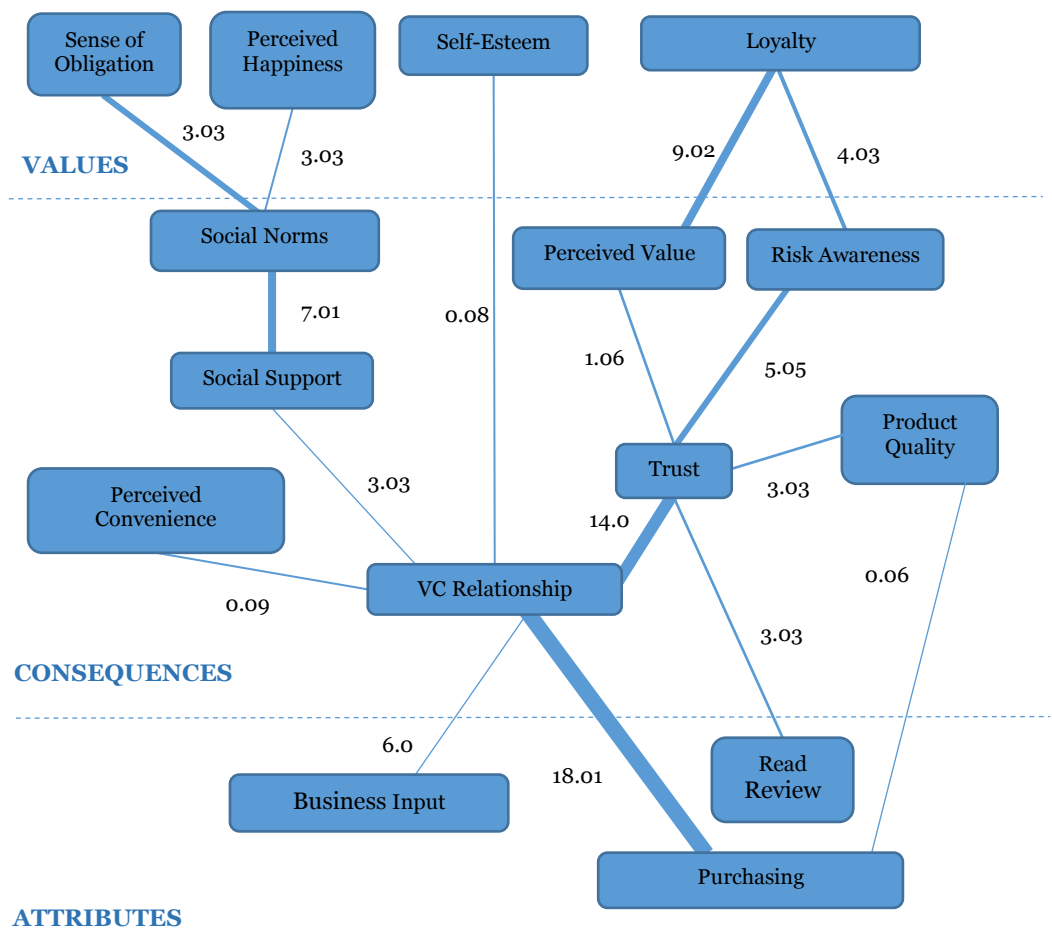

Figure 1: Hierarchical Value Map

## 4  Results and discussion

This study found that the main social commerce activities carried out by customers in the community are (i) purchasing, (ii) reading reviews and (iii) business input. Business input activities include actions that can increase the worth of goods, services or the businesses. The HVM also shows that the main goals for engaging in social commerce are (i) loyalty, (ii) self-esteem, (iii) perceived happiness, and (iv) sense of obligation.

This study will focus on the strongest basic motivation which is the VC relationship. VC relationship was mentioned 28 times in 46 ladders, making it the main reason why community members wanted to purchase from sellers in social commerce settings as compared to other factors such as trust and product quality. Virtual community (VC) relationship is defined as the personal friendships developed among members in blogosphere community. These relationships often develop through private online communication, and they sometimes move into face-to-face interactions around the common topic of interest. The relationship is based on previous interactions, experiences, and learning within the virtual community environment.

The relationship between purchasing activities and VC relationship denotes 18.01 linkages, indicating that purchasing activities are strongly influenced by the relationships within the community. It shows that 18 purchasing activities from 21 participants are directly because of the VC relationship. Business input activities are also strongly influenced by VC relationship as shown by 6.0 linkages. Six is relatively low, however compared to the number of times it was mentioned in this study (n=10), it shows that more than half the participants who participated in business input activities were influenced by their relationship to the community. However, no direct linkage was found between reading review activities and VC relationship, therefore it was dropped from the discussion.

In order to understand how higher level elements on the HVM that are connected to the VC relationship, and what are the corresponding attributes associated with VC relationship, two chains (A to V) were extracted from the HVM, namely (i) VC relationships –> trust path, and (ii) VC relationships –> social support path.

### 4.1  VC relationship –> trust path

Figure 2 presents the extracted HVM for VC relationship –> trust path. The attributes or activities involved in this path include purchasing and business input. As shown in Figure 2, VC relationship can lead directly to trust. The direct relationship from VC relationship is 14 which is relatively high. The fact that the VC relationship only has a direct relationship but no indirect linkages with trust shows that that the relationships within the community are crucial to the development of trust. The ffindings show that customers indicate that they need to trust the sellers in the community before they are motivated to participate.

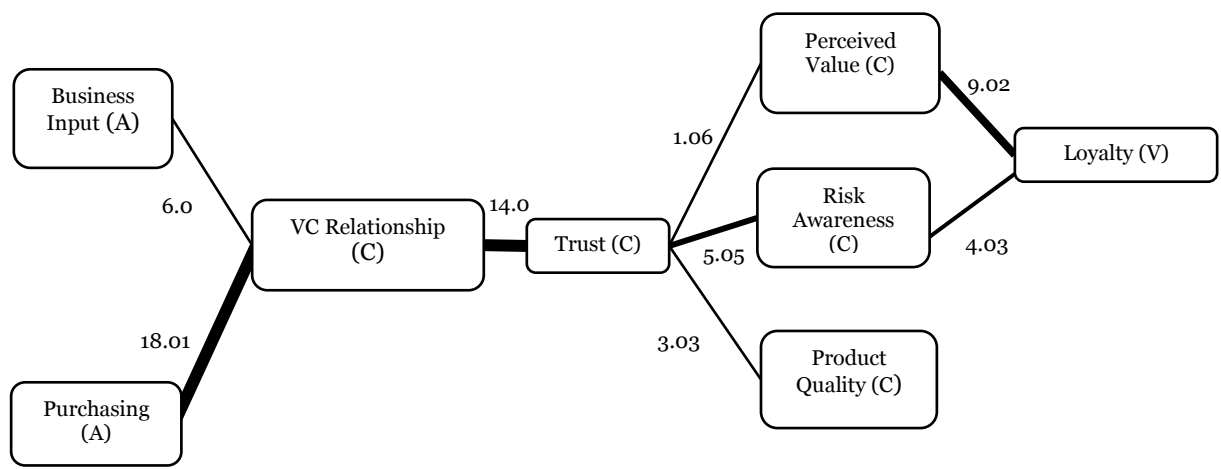

Figure 2: VC relationship –> trust path

This findings supports some of the findings in previous work conducted by Hsu, Liu and Lee (2010) on relationship marketing. Hsu et al. (2010) found that trust in enterprise micro-blogs could enhance consumers' commitment and sense of community. However, Hsu et al. (2010)'s findings are limited to the enterprise business setting in Taiwan and in that case trust was developed using the enterprise's strong corporate image. Our finding show that the trust developed in lifestyle blogging community is created not only from what customers read on blogs, or from a corporate image but is mainly influenced by the personal relationships that readers have with the bloggers in their daily communications within the community, this is developed long before the bloggers venture into social commerce.

It is apparent in Figure 2 that trust can lead to perceived value and risk awareness. Risk awareness refers to a customer's awareness of the uncertainty and adverse consequences of engaging in an activity. The trust that develops in the community lowers the perception of risks by the customers. Meanwhile perceived value refers to customers' perception of the complete shopping experience, not only product acquisition but also including the relativistic preference that can characterize a customer's experience of interacting with some object. This includes both tangible and intangible benefits gained from the purchase or participation in community. The findings show that the value perceived by the customers includes the following (i) extrinsic/utilitarian value, (ii) economic value, (iii) hedonic value and (iv) epistemic value (Babin et al. 1994, Holbrook 1994, Holbrook 1999, Lee and Overby 2004, Mattsson 1991, Smithand Colgate 2007, Sweeney and Soutar 2001, Sweeney et al. 1999).

Extrinsic/Utilitarian value is instrumental, task-related, rational, functional, cognitive, and a means to an end; or an overall assessment of functional benefits incorporating the following dimensions: traditional price saving; service; time-saving; and merchandise selection. Economic value represents a measure of the benefits provided by a good or service to an economic agent. It is generally measured relative to units of currency, and the interpretation is therefore "what is the maximum amount of money a specific actor is willing and able to pay for the good or service"?. Hedonic value reflects the entertainment and emotional worth of shopping; non instrumental, experiential, and affective. It is also concerned with the extent to which a product creates appropriate experiences, feelings, and emotions for the customer. Lastly epistemic value is concerned with a desire for knowledge which can be motivated by intellectual curiosity or the seeking of novelty.

This study found that customers acknowledge the value that they will receive (perceived value) as well as acknowledging the risk (risk awareness) they might have if there is no trust in their social commerce participation. However, for trust –> the perceived value path, compared to direct linkage, the indirect linkages were stronger (1.06). This means there are unknown factors that lead from trust to perceived value. Although the relationship between trust and perceived value has been observed to promote loyalty in e-commerce by previous researchers (Petrick and Backman 2001, Yang et al. 2004, Gallarza et al. 2006, Sirdeshmukh 2008, Jin 2008), nonetheless we found that the value perceived from trust in this study was not only related to product acquisition but also to overall experience.

Meanwhile for trust – the direct and indirect linkages for the risk awareness path were 5.05 which was considered moderate as risk awareness was mentioned 11 times in the customers' section. This supports the findings in Kim, Ferrin and Rao (2008) who found that trust to a large degree addresses the risk problem in e-commerce in two ways: by reducing risk and by increasing purchase intentions directly.

The results also show that risk awareness and perceived value have a direct and indirect relationship with the main reason why customers participate in social commerce, which is loyalty. Customers' perceived value has a stronger relationship with loyalty at 9.04 whilst risk awareness to loyalty is 4.02. This shows that customers acknowledge that developing trust in their virtual relationship will lead to a higher element in the HVM, which is the loyalty to the sellers.

The last element connected in VC relationship –> trust path is the product quality. Although the linkage is quite low at 3.03, it shows that customers realize the important of relationship and trust in the community in order to ensure a good quality product. This finding supports existing theories in e-commerce research, such as Kim et al. (2009) and Wong and Sohal (2002).

### 4.2       VC Relationship –> Social Support Path

Figure 3 shows the VC relationship –> social support path. Again as in the previous path, both purchasing and business input are involved. Social support in this research refers to a members' experiences of being cared for, being responded to, and being helped by other blogosphere community members (and vice-versa). The linkages between VC relationship and social norms are shown to be moderate with direct and indirect linkages of 3.03 as compared to the VC relationship and trust path. This shows that the relationships between the community members, particularly the customer – seller relationship is likely to create social support within the community. There are also some unknown indirect influences from this relationship that can lead to social support.

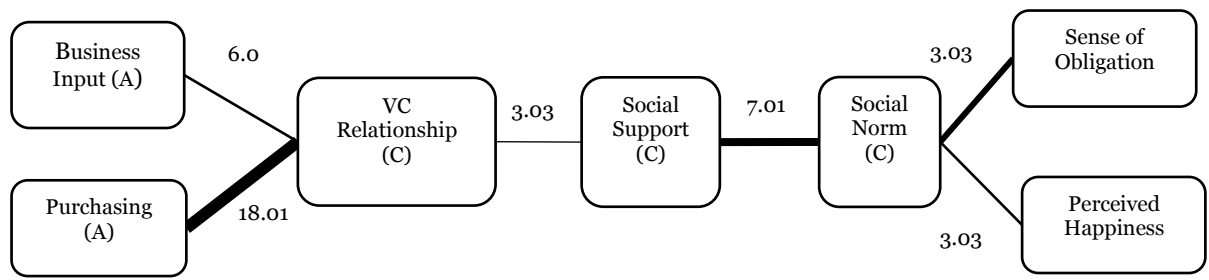

Figure 3: VC relationship –> Social support path

The findings also show that the customers acknowledge that social support is not only limited to personal support but also to business related support. Other researchers (Hajli 2014) have also found that social support with emotional and informational dimensions is being created in Social Networking Sites (SNS), and refers to the concept as online social support. According to Hajli (2014), this factor is a key construct in the development of e-commerce into social commerce. Mostly because supportive interactions among consumers in SNS produces stronger relationships, making them feel closer to peers. Although this is similar to our findings, nonetheless, his study only focuses on relationship building in Facebook. In addition, we also discovered the existence of four further types of social support between customers and sellers, including, emotional support, instrumental support, informational support and appraisal support. According to Barrera and Ainlay (1983), Sarason et al. (1987), Barrera Jr (1986), Wills (1985) and Uchino (2004), the four types of social support are defined as follows: (i) Emotional Support- expressions of empathy, love, trust and caring; (ii) Instrumental Support - tangible aid and service; (iii) Informational Support- advice, suggestions, and information, (iv) Appraisal – provides information that is useful for self-evaluation.

In addition, awareness about the practice of giving social support is one of the reason why customers provide social support to others within the community, particularly to the bloggers. This explains the next element in this path, social norms. Social norms are defined as the appropriate behaviour of members in the blogosphere community or the pattern of behaviour in the community, or culture, which is accepted as normal and to which a member is expected to conform. The direct relationship from social support to social norms is 7 which is stronger than the indirect relationship (1). The findings indicate that customers are likely to provide social support because everybody else also does the same thing. This shows how the strong sense of virtual community can influence the customers to participate in social commerce.

The upper elements of social norms are (i) sense of obligation and (ii) perceived happiness, which are the values or motivations for customers to participate in social commerce for this path. The direct and indirect linkages between social norms and both elements are 3.03. The social norms and sense of obligation path will be discussed first. The sense of obligation refers to members' commitments, which is evident in the desire to maintain community relationships. The direct linkage from social norms to sense of obligation is equal to the indirect linkage (3.03), which denotes a weak linkage. However, if compared to the total number of times sense of obligation is mentioned in this study (n=10), this can be considered a moderate relationship. The reason why customers feel that they need to conform to these social norms is because they feel obligated to the community. The reasons include (i) nationalism, (ii) benevolence and (iii) sense of responsibility. The nationalism in this matter includes purchasing because of the same culture as well as purchasing to support local products.

Next, the relationship between social norms with perceived happiness will be discussed. Perceived happiness refers to the state of feeling associated with the members' well-being. The direct and indirect linkages are 3.03 which means that social norms being the reason for customers' happiness is equal to several unknown factors in between the relationships. This finding indicates that social norms lead to the happiness and satisfaction of the customers in the community.

This path basically supports the important role of virtual community as covered in the literature by Koh and Kim (2003), Blanchard and Marcus (2004), and Ellonen et al. (2008). It shows that members in the community are participating in social commerce because they wanted to give back to the community.

# 5 Conclusion

The laddering technique explores the customers' motivation by connecting the relationships between constructs into the HVM. It shows how one factor relates to other factors in order to understand the motivation behind social commerce activities in this community. The use of the laddering method has uncovered a full range of attributes, consequences and values associated with the customers' decisions to participate in social commerce, which provides insights into customers' core beliefs (Hawley 2009). Using laddering techniques has given a deeper understanding of this problem. Some scholars have highlighted issues concerning the loss of data when using the laddering technique (Saaka et al. 2004, Baker 2002) during the transformation of the implication matrix to HVM. Nonetheless we argue that the missing data in some chains shows that the factors are not shared by majority of the participants. This shows that the factors can't be considered as the main influential factors in the chain as they are not common factors amongst the customers in this community. Therefore we argue that the loss of data is not a significant issue in our study.

The results indicate that the main factors that influence the customers are underpinned by the strong sense of virtual community among community members. This is central to understanding virtual communities and members' participation in social commerce. This is crucial for the sustainability of social commerce in a small non-brand communities. It is hoped that the findings will help researchers addressing various academic issues related to social commerce implementation, not only to provide a better understanding of the phenomenon, but also to stimulate reflections on its current stage and its future direction.